\DocumentMetadata{}
\documentclass[sigconf,nonacm,acmsmall,natbib=false]{acmart}
\usepackage{graphicx} % Required for inserting images
\usepackage[frozencache,cachedir=./mintcache]{minted}
\usepackage{xcolor}
\usepackage[%
    natbib=true,
    style=numeric,
    sorting=none,
    backend=biber%
]{biblatex}
\usepackage{siunitx}
\usepackage{todonotes}
\usepackage{hyperref}
\usepackage{subcaption}

\newmintinline[bash]{bash}{breaklines=true,fontsize=\footnotesize,autogobble=true}
\title[Best Practices for HPC LLM Training]{Training LLMs on HPC Systems: Best Practices from the OpenGPT-X Project}
\author{Carolin Penke}
\email{c.penke@fz-juelich.de}
\author{Chelsea Maria John}
\email{c.john@fz-juelich.de}
\author{Jan Ebert}
\email{ja.ebert@fz-juelich.de}
\author{Stefan Kesselheim}
\email{s.kesselheim@fz-juelich.de}
\author{\\Andreas Herten}
\email{a.herten@fz-juelich.de}
\affiliation{%
    \department{Jülich Supercomputing Centre}
    \institution{Forschungszentrum Jülich}
    \city{Jülich}
    \country{Germany}%
}

\date{February 2025}
\begin{document}
% Can we add an abstract in this style?
\maketitle
\begin{center}
    \textit{March 31, 2025}
\end{center}

\section{Introduction}
The field of Artificial Intelligence (AI) has experienced a rapid boost with the advent of performant Large Language Models (LLMs). These have transformed Natural Language Processing (NLP), driving advancements in machine translation, text summarization, and conversational AI. Training these models requires vast computational resources, necessitating the use of High Performance Computing (HPC) systems. Within the OpenGPT-X\footnote{\url{https://opengpt-x.de/en/}} project, an initiative focused on developing open and sovereign European AI language models, computational resources required for training  were provided by the supercomputer JUWELS Booster~\cite{krause2019juwels} at the Jülich Supercomputing Centre (JSC), facilitating a scalable and optimized LLM training.

A key milestone of the OpenGPT-X project is the release of Teuken-7B~\cite{ali2024teuken7b}, a 7-billion-parameter model designed to support multiple European languages with an emphasis on transparency, efficiency, and open access. Training Teuken-7B on JUWELS Booster leveraged the supercomputer’s GPU-accelerated architecture to efficiently manage large-scale data processing, optimize memory usage, and ensure high training throughput. The model's development required advanced parallelization strategies, including data parallelism, pipeline, tensor and sequence parallelism, as well as mixed-precision training to fully utilize JUWELS Booster’s computational power.

In this work, we detail the methodologies employed for training Teuken-7B on JUWELS Booster, addressing challenges related to scalability, resource efficiency, and model convergence. We also present a benchmarking framework and insights gained from this large-scale training effort.

\section{Background}
In this section, we provide a concise overview of key developments in the field of Natural Language Processing (NLP) during the duration of the OpenGPT-X project (2022–2025), emphasizing the role of HPC centers as critical partners. The contents of this section support the recommendation to closely follow the rapid advancements in NLP and AI, while keeping project goals focused. For broader takeaways and lessons learned, see \autoref{Sec:Conclusions}.

\subsection{Large Language Models}
Given a sequence of text tokens, a \emph{language model} predicts the next most-probable token and can be used for autonomous text generation in a variety of tasks. 
Most modern language models are deep neural networks based on the transformer architecture published in 2017~\cite{vaswani2023attentionneed}, and are trained on vast amounts of text data. Previously, variants of recurrent neural networks (RNNs), such as long-short-term memory (LSTM) networks dominated the field of sequence learning. The attention mechanism, originally proposed as a feature of RNNs, is the main building block of the transformer architecture. This architecture avoids recurrence and convolutions, and thus training and inference can be highly parallelized. By mainly relying on matrix-matrix products, operations can be performed in an efficient manner on modern hardware such as GPUs~\cite{penke2022transformer}. 

In 2018, OpenAI published GPT (later to be known as GPT-1), a language model based on a simplified decoder-only transformer architecture, with an increased number of layers, using 117 million parameters. It introduced the paradigm of unlabeled pre-training on large corpora of text, followed by task-specific fine-tuning. The BERT model ~\cite{devlin2018bert} (up to 340 million parameters) was published by Google a few months later and employed a masked bidirectional pre-training objective, making it suitable as a text encoder. GPT-2~\cite{radford2019gpt2} followed in 2019 and was based on the transformer architecture with slight modifications. It further increased the number of parameters to 1.5 billion. With its increased capabilities, it paved the way to move from task-specific models trained on labeled data, towards general models, that have been trained unsupervised. With GPT-3~\cite{brown2020gpt3} in 2020, the number of parameters rose to 175 billion, and once more a drastic jump in capabilities was observed. 

Under this backdrop, the OpenGPT-X project was devised. During the project runtime, InstructGPT~\cite{ouyang2022instructgpt} applied Reinforcement Learning from Human Feedback (RLHF)~\cite{ziegler2020rlhf,stiennon2022rlhf} in order to turn the pre-trained text completion models into helpful chat assistants, aligned with human preferences. Based on InstructGPT, ChatGPT was published in 2022, making the impressive capabilities available to the general public in an intuitive user interface. 

Further models by OpenAI, such as GPT-4 followed and a number of commercial competitors started to offer similar systems, among them Google with Gemini, Anthropic with Claude, and Groq. Due to commercial interests, most of these models, specifically their pre-trained weights, are not publicly available (closed source). The level of detail in published specifications varies. For example, the number of parameters of GPT-4 is not disclosed, but multiple sources have stated GPT-4 to be based on a mixture-of-experts model employing 1.8 trillion parameters\footnote{\url{https://the-decoder.com/gpt-4-architecture-datasets-costs-and-more-leaked/}, accessed 2025-03-31}. 

Further ChatGPT competitors arose in the open-source space. Meta published a series of Llama models~\cite{touvron2023llama}. Most recently, Deepseek~\cite{deepseekai2024} published an open-weights model with 671 billion parameters on par with the capabilities of recent OpenAI models. 

\subsection{OpenGPT-X}
OpenGPT-X is a German collaborative initiative aimed at developing large-scale, multilingual language models tailored to business and research needs. It emphasizes transparency, trustworthiness, and open-source principles. Funded by the German Federal Ministry for Economic Affairs and Climate Action (BMWK), the project spanned from January 2022 to March 2025 and released Teuken-7B, a 7-billion-parameter model optimized for European languages, along with several research publications~\cite{penke2022ogxposter,ostendorff2023efficientlanguagemodeltraining, john2023poster, ali2024tokenizerchoice,john2023novel,weber2024multilingual,mirza2024illuminer,courtois2024symmetric,ostendorff2024llmdatasets,john2024Caraml,ali2024teuken7b,thellmann2024multilingual,brandizzi2024data, ebert2025nic}.

The initiative is part of the broader Gaia-X ecosystem, a European project dedicated to creating a secure, transparent, and sovereign data infrastructure. Gaia-X fosters collaboration across various industries, supporting digital sovereignty and innovation by establishing common standards and trusted data spaces for sharing data securely and transparently across Europe.

OpenGPT-X addresses multiple aspects necessary for developing and deploying advanced language technologies. It encompasses utilizing scalable GPU-based computing infrastructures, creating compatible data environments and MLOps workflows, developing innovative AI language services, and ensures interoperability with established AI platforms. Additionally, it focuses on real-world prototyping and application testing across diverse domains thereby validating practical applicability for industry and research.

The focus of this report is the establishment of a highly scalable GPU-based computing infrastructure, central to the OpenGPT-X project. By integrating the computational resources from key consortium partners, including Fraunhofer, IONOS, Forschungszentrum Jülich, and TU Dresden, the project has created an environment that supports large-scale training of AI language models. Furthermore, this infrastructure has served as a testing ground for evaluating and optimizing emerging GPU and accelerator technologies, reinforcing Europe's technological sovereignty in AI.

\subsection{Current Trends and Developments}

In this subsection, we outline and briefly describe significant trends in language model research and development that emerged during the OpenGPT-X project and are expected to continue influencing the field.

\begin{enumerate}
    \item \textbf{Number of parameters:} 
    While leading pre-trained flagship models have grown significantly beyond GPT-3 in terms of parameter count, it has become apparent that sheer scale alone is insufficient to address fundamental limitations of LLMs. The availability of data to saturate large parameter counts has turned out to be a significant bottleneck~\cite{hoffmann2022chinchilla}. In contrast, smaller, more efficient models have gained relevance, especially in industrial applications, as they are easier to host with limited hardware capabilities. 
    
    \item \textbf{Transformer architecture developments:} Improvements include novel positional embeddings, such as Rotary Positional Embedding (RoPE)~\cite{su2023rope}, Mixture-of-Experts models~\cite{jiang2024mixtralexperts,muennighoff2025olmoe}, attention variants such as grouped-query attention (GQA)~\cite{ainslie2023gqa} and multi-head latent attention (MHLA)~\cite{deepseekai2024}.
    
    \item \textbf{Parallelization and other computational optimizations in training:} In terms of training parallelization techniques, fully-sharded data parallelism (FSDP)~\cite{zhang2022opt,zhao23fsdp,wang2024memorybandwidthfsdp} was further established as an alternative to 3D parallelism with ZeRO memory optimizations~\cite{rajbhandari2020zero}. Sequence parallelism was developed as an additional level of parallelism, often combined with activation recomputation~\cite{korthikanti2022reducingactivation}.   Hardware-aware  techniques such as Flash-Attention~\cite{dao2022flashattention,dao2023flashattention2}, and evolved applications of mixed-precision~\cite{deepseekai2024} have significantly increased training and inference efficiency.
    
    \item \textbf{Optimizations outside of pre-training:} Advances have occurred on the whole life cycle of model training. These include data preprocessing~\cite{brandizzi2024data}, tokenization schemes~\cite{ali2024tokenizerchoice}, hyperparameter optimization (e.g., Maximal Update
Parametrization~\cite{yao2023mup}), fine-tuning approaches such as reinforcement learning with human feedback (RLHF)~\cite{ziegler2020rlhf, stiennon2022rlhf}, and methods to enhance reasoning capabilities, including prompt engineering~\cite{deepseekai2024}. Inference efficiency is improved with techniques such as  KV-Caching~\cite{pope2022kvcache}.
    
    \item \textbf{Emergence of novel architectures:} Beyond traditional transformers, alternative neural network architectures such as state-space models (SSMs) are gaining attention~\cite{gu2024mamba}, promising improvements in performance and computational efficiency. Another promising approach is explored by xLSTMs~\cite{beck2024xlstm}.
    
    \item \textbf{More powerful hardware:} Continual advancements in GPU and accelerator technologies have supported faster model training, increased scalability, and reduced costs. One example of AI workloads shaping chip designs are tensor cores and low-precision support, e.g. exploited by NVIDIA's Transformer Engine library. 
    
    \item \textbf{Enhanced usability and system integration:} Interfaces like ChatGPT’s conversational assistant have significantly lowered entry barriers, making powerful language models accessible to broader audiences. This category includes agentic frameworks, , where models interface dynamically with external tools, APIs, and environments to expand their applicability.
    
    \item \textbf{Multimodal capabilities:} Multimodal models, such as those based on Vision Transformers~\cite{dosovitskiy2021vision}, become more common. They integrate text, images, audio, and video to support a wider range of applications.
\end{enumerate}

\section{Hardware Overview}\label{Sec:Hardware}
Along with the Center for Information Services and High Performance Computing (ZIH) at TU Dresden, the Jülich Supercomputing Centre provided the main computational resources required to train large language models within OpenGPT-X. In this section we give details about the used systems at the Jülich site. The JUWELS supercomputer employs the Modular Supercomputing Architecture~\cite{suarez19MSA}. Two of its modules, the CPU-based JUWELS Cluster and the GPU-accelerated JUWELS Booster, were used within the OpenGPT-X project. 

To secure computing resources annually, we applied to the Calls for Large-Scale Projects issued by the Gauss Centre for Supercomputing (GCS). Our proposals underwent rigorous peer review, and compute time was allocated through a competitive selection process.

\subsection{JUWELS Booster}
Training large language models is highly computationally demanding and requires state-of-the-art GPUs. Within OpenGPT-X, the training was performed on the JUWELS Booster module, which employs NVIDIA A100 GPUs and is designed to handle computationally intensive tasks, such as large-scale simulations and AI workloads. This module provides the bulk of the JUWELS supercomputer’s floating point operation capacity (FLOP/s), making it one of the fastest supercomputers in the world. During the OpenGPT-X project it ranked 33rd globally, while it held the 7th position at the time of its inception.

The Booster module consists of 936 compute nodes, each equipped with:
\begin{description}
    \item [CPUs] Two AMD EPYC Rome 7402 processors, totaling 48 cores per node, running at 2.8 GHz.
    \item [Memory] 512 GB DDR4-3200 RAM.
    \item [GPUs] Four NVIDIA A100 Tensor Core GPUs with 40 GB HBM2e memory each, interconnected via NVLink3.
    \item [Network] Four Mellanox HDR-200 InfiniBand adapters per node, enabling high-speed data transfer.
\end{description}

Booster uses a DragonFly+ network topology to connect its compute nodes via HDR-200 InfiniBand. Here, the nodes are organized into 20 cells, each containing 48 nodes connected in a full fat-tree topology. This design ensures efficient intra-cell communication with a bi-section bandwidth of 40 Tbit/s. Pairs of cells are connected with 10 links each, providing 4 Tbit/s bisection bandwidth between cells and a total of 400 Tbit/s bisection bandwidth across the system.

\subsection{JUWELS Cluster}
The OpenGPT-X project was also granted compute time on the JUWELS Cluster module. The Cluster module contains mostly standard compute nodes, which employ two Intel Xeon Platinum 8168 CPUs with 24 cores each, running at 2.7 GHz. The accelerated compute nodes were also of interest, as they contain four V100 GPUS per node, representing an earlier GPU generation than the Booster module. JUWELS Cluster was used for various data preprocessing and experimentation tasks~\cite{brandizzi2024data}. 

\subsection{Other Platforms}
\label{Sec:JurEval}
Extensive testing and benchmarking (find details in \autoref{Sec:Benchmark}) were performed on several other systems in order to observe the market and evaluate promising new technologies. The JURECA Evaluation Platform~\footnote{\url{https://apps.fz-juelich.de/jsc/hps/jureca/evaluation-platform-overview.html}} was essential in this endeavor. This platform is an extension of the JURECA Cluster~\cite{Thornig2021jureca}, incorporating individual nodes containing accelerators from various vendors. For example, the first device of the new NVIDIA H100 GPU generation was installed here for early testing, before it found its way into newer HPC systems. Specifically, the following systems were used.
\begin{description}
    \item [NVIDIA H100] One node, equipped with four H100 GPUs (PCIe), containing 80 GB of memory each.
    \item [NVIDIA GH200] Two nodes, containing one NVIDIA Grace Hopper Superchip, with 480GB CPU memory, and 96GB of GPU memory. 
    \item [AMD MI200] One node, equipped with four MI250 GPUs, containing 128 GB of memory each
    \item [Graphcore IPU-POD4] Equipped with four GC200 IPUs. 
\end{description}
The Graphcore node is the most unconventional one, as it does not use a typical GPU architecture, but represents a dataflow approach. The cores are equipped with local memory and are specialized to run graph-based computations, such as neural network training, efficiently, employing a powerful on-chip network. It represents a new class of promising AI accelerators, put forward by companies such as Cerebras, Groq, SambaNova and Tenstorrent.

The GH200 superchip distinguishes itself from a regular CPU with an attached H100 GPU, because here, CPU and GPU are tightly integrated. A high-speed interface allows for a coherent memory model.

As the OpenGPT-X project progressed, at JSC the new JEDI development system\footnote{\url{https://www.fz-juelich.de/en/ias/jsc/systems/supercomputers/jedi}} was deployed in 2024. JEDI represents a first module of the exascale supercomputer JUPITER, to be deployed in 2025. Each of the 48 compute nodes contains 4 NVIDIA GH200 Grace-Hopper superchips with 120GB CPU memory and 96GB GPU memory each.

Resources from the WestAI consortium\footnote{\url{https://westai.de/}} were also employed to evaluate differences in specific H100 node setups~\cite{john2024Caraml}.

\section{Software}\label{Sec:Software}
The primary objective of the OpenGPT-X project was to train deep neural networks with a transformer architecture~\cite{vaswani2023attentionneed} on language data to create language models that can be used for text generation. 
Established deep learning frameworks such as PyTorch~\cite{paszke2019pytorch} and TensorFlow~\cite{abadi2015tensorflow} allow specifying neural network architectures and managing the training process effectively. As a backend, the highly performant NCCL library is employed to enable efficient collective communications on large-scale GPU clusters.  

A \emph{training codebase} describes a collection of scripts and tools and contains the main training loop as a core functionality. Here, optimizer steps are carried out using backpropagation based on the specified architecture. Parallelization strategies and optimizations can be implemented and helper functions are provided, e.g. for applying a tokenizer in a preprocessing step, or for checkpoint conversion.  

\subsection{Choosing a Training Codebase}
At the beginning of the OpenGPT-X project, several well-maintained open-source codebases already existed. Developing entirely new software from scratch would have been an inefficient use of resources. Instead, the project focused on adapting and customizing existing solutions to its specific needs.

The OpenGPT-X project initially adopted the Megatron-DeepSpeed codebase~\footnote{\url{https://github.com/bigscience-workshop/Megatron-DeepSpeed}}, developed by NVIDIA, extended by Microsoft researchers and further adapted during the BigScience research workshop~\cite{workshop2023bloom176b}. Other codebases, such as Meta’s Open Pretrained Transformer (OPT)~\cite{zhang2022opt}, also emerged, promising potential advantages in abstraction and usability. After evaluating multiple codebases, BigScience’s implementation was chosen for its superior community support, adaptability, and better documentation compared to OPT. BigScience's integration of Megatron-DeepSpeed with NVIDIA’s Megatron-LM framework offered 3D parallelism and ZeRO memory optimizations~\cite{rajbhandari2020zero} as advanced parallelism techniques which had proven effective for large-scale language model training. OPT's parallelization strategy, based on Meta’s Fully Sharded Data Parallel (FSDP), showed lower efficiency at scale, and potential licensing issues further reinforced the decision to adopt BigScience’s solution.

As the project progressed, the independent Megatron-LM~\footnote{\url{https://github.com/NVIDIA/Megatron-LM}} repository, NVIDIA's flagship codebase, emerged as a more actively maintained option. Consequently, OpenGPT-X transitioned to Megatron-LM to leverage NVIDIA’s GPU kernel optimizations, resulting in greater training stability, increased efficiency, and a 75\% reduction in iteration time. Megatron-LM also allowed us to streamline the environment setup.  A smooth knowledge transfer from the previously used Megatron-DeepSpeed was ensured. 

The project incorporated state-of-the-art techniques such as Flash-Attention 2~\cite{dao2023flashattention2}, sequence parallelism~\cite{korthikanti2022reducingactivation}, ALiBi embeddings~\cite{press2022alibi}, and the Adan optimizer~\cite{xie2023adan}, alongside innovations in hyperparameter scaling~\cite{yao2023mup}, sparse matrix optimization, and GPU kernel improvements. Collectively, these enhancements resulted in substantial performance gains and improved overall efficiency~\cite{ebert2025nic}.

To support researchers and engineers in understanding the complexity of parallelism and scaling in large language model training, resources such as the UltraScale Playbook~\cite{ultrascaleplaybook} offer an excellent summary and interactive tools for exploring these concepts.

\subsection{The Setup Repository}
To deploy the training codebase within specific computing environments of multiple HPC clusters, additional adaptation and integration efforts were required. This involved developing customized integration code, which we consolidated into a dedicated repository known as the \emph{setup repository}. This repository contains site-specific documentation, detailed instructions, and example configurations necessary to initiate training runs. At Jülich Supercompuuting Centre (JSC), resource management and job scheduling are handled using Slurm, a widely adopted workload manager in HPC environments. Slurm scripts specify computational resource requirements and orchestrate job execution on the cluster. Special care was taken during development to ensure compatibility with the existing system software stack, primarily provided through EasyBuild installations and managed via environment modules~\cite{geimer14LmodEasybuild}. Initially, Python virtual environments were used to manage Python package dependencies, complemented by environment modules for external software. Later, container-based approaches leveraging Apptainer~\cite{singularity21} were introduced, customized, and documented for seamless integration into the specific system environments. See \autoref{Sec:Training} for details on the training process.

The setup repository and computational environments for training LLMs on HPC systems were continuously maintained and regularly updated with the latest software packages and necessary patches. To ensure seamless operation across both participating HPC centers (Jülich and Dresden with clusters Taurus and Alpha Centauri), their respective setup repositories were unified, enabling users to easily transition training workflows between these environments. Beyond optimized training scripts, data preprocessing utilities were developed and maintained to efficiently convert data into the binary format required by the codebase and models. Ultimately, the setup repository provided a robust foundation for integrating the training workflows into a continuous benchmarking pipeline within the JUPITER Research and Early Access Program (JUREAP).

\section{Data Storage}
The JSC provides a robust file system infrastructure on its supercomputing platforms through a multi-cluster General Parallel File System (GPFS), centrally managed by the Jülich Storage Cluster (JUST)~\footnote{\url{https://apps.fz-juelich.de/jsc/hps/just/configuration.html}}. Users have access to three distinct storage partitions, \bash{$HOME}, \bash{$PROJECT}, and \bash{$SCRATCH}, for managing and processing data. Within the OpenGPT-X project, approximately 800 TB of \bash{$SCRATCH} storage and 15 TB of \bash{$PROJECT} storage were used, organized according to a structured data-flow design.

One challenge related to data storage involved frequently running into storage capacity limitations, including both physical storage space and inode exhaustion. To address this, users were supported in their efforts to clean up unused data while we successfully applied for additional storage capacity where needed. Special care was required for shared temporary directories used during setup and training. Additionally, default cache directories posed a challenge, as they often led to extensive cache data being written to suboptimal storage partitions.

For data movement between file system or as a gateway to external systems, the JUDAC (Jülich Data Access) server was used. Given the involvement of two separate HPC centers in the project, efficient data transfer mechanisms became crucial. To facilitate this, administrators from JSC and TU Dresden (TUD) collaborated to establish UNICORE FTP (UFTP)~\cite{uftp}, enabling seamless, high-performance transfers of large-scale datasets across sites.

Additional storage systems based on non-volatile memory technologies (NVMe) were also evaluated during the project. Notably, the High-Performance Storage Tier (CSCRATCH) was assessed as a user-managed caching layer built on top of the existing \bash{$SCRATCH} filesystem. Furthermore, through an industrial collaboration with VAST Data, a proof-of-concept use case was developed to test innovative storage solutions. The NVMe-based storage systems demonstrated significant performance improvements, particularly for checkpointing. Here, all parameters and optimizer states are saved to persistent storage after a predefined number of training steps. This enables efficient recovery and continuation of training in case of undesirable training developments or interruptions due to hardware failures or job time constraints.

\section{Training}\label{Sec:Training}
Effectively utilizing HPC systems can pose significant challenges, particularly for users with diverse technical backgrounds. To ensure a smooth onboarding process, tailored guidelines and detailed introductions were provided via the project's centralized communication platform, Confluence. Additional comprehensive documentation is available in the official JSC documentation\footnote{\url{https://apps.fz-juelich.de/jsc/hps/juwels/index.html}}.

To access the clusters, users must first obtain a JSC account through the JuDoor portal\footnote{\url{https://judoor.fz-juelich.de/}}. These accounts are then associated with specific compute time projects that correspond to the compute resources awarded by the GCS call (see \autoref{Sec:Hardware}). Users connect to the HPC clusters via SSH, initially accessing a login gateway node. From this node, computational jobs -specified in the form of job scripts- are submitted to the Slurm batch scheduler using the \bash{sbatch} command. A detailed overview of the software environment and infrastructure used is provided in \autoref{Sec:Software}. An example job script for running LLM training on JUWELS Booster using Megatron-LM and NVIDIA PyTorch containers is available in \autoref{LLMSbatch}.

As illustrated in the appendix, configuring functional job scripts for large-scale trainings is a non-trivial task. Significant effort went into resolving the intricacies involved in running PyTorch effectively on massively parallel HPC clusters. In the following, we highlight some of these specific challenges, referring readers to the comprehensive "PyTorch at JSC" guide\footnote{\url{https://gitlab.jsc.fz-juelich.de/sdlaml/pytorch-at-jsc}} for further information.

In order to diagnose bugs and treat them effectively, enabling higher levels of debug logging was essential. Environment variables used in this regard included \bash{$NCCL_DEBUG}, \bash{$NCCL_ASYNC_ERROR_HANDLING} (now \bash{$TORCH_NCCL_ASYNC_ERROR_HANDLING}), \bash{$TORCH_DISTRIBUTED_DEBUG}, and \bash{$CUDA_LAUNCH_BLOCKING}.

\subsection{Hardware Challenges}
During the OpenGPT-X project, a hardware-related issue known as "link-flipping" was identified with the InfiniBand cabling on the Atos Sequana XH2000 systems at JSC. This issue predominantly affected large-scale training jobs and proved difficult to reproduce at smaller scales. Link-flipping refers to the InfiniBand adapter temporarily losing its link, causing delays in reconnection. It can lead to intermittent failures, especially with the NCCL library and PyTorch. The problem was attenuated by retrofitting InfiniBand cables with ferrite beads and modifying certain environment variables, such as \bash{$NCCL_IB_TIMEOUT}, \bash{$UCX_RC_TIMEOUT}, and \bash{$NCCL_IB_RETRY_CNT}. 

Additionally, the job scripts used for training were modified to trigger immediate checkpointing when Slurm encountered errors. Combined with regularly scheduled checkpointing at predefined intervals, these modifications ensured that training could be seamlessly resumed after interruptions caused by hardware failures or other unforeseen issues.

\subsection{Software Challenges}
This subsection provides examples of software-related challenges encountered in the given HPC environments. It is intended as an illustrative rather than an exhaustive list.

One aspect to consider is the pinning behavior of processes and threads, i.e., exactly how they are distributed to available CPU cores. Pinning can be managed at various levels: Typically, a system-wide default configuration exists, but users may override this via settings provided by the job scheduler. For JSC systems, detailed instructions can be found in the official user documentation\footnote{\url{https://apps.fz-juelich.de/jsc/hps/juwels/affinity.html}}. With Slurm, process pinning can be controlled using the \bash{--cpu-bind} flag within the \bash{srun} command in a job script. \autoref{LLMSbatch} shows the use of \bash{--cpu-bind=v}, which provides verbose output detailing the binding employed on the job level. In certain scenarios when observing a low performance, explicitly disabling Slurm’s pinning with \bash{--cpu-bind=none} and delegating the optimization of process placement to specialized software, such as the NCCL communication library, may yield better results.

Using containers on HPC systems introduces unique challenges compared to simpler workstation setups. While containers offer encapsulation, isolating the software from external environmental dependencies, this isolation must be balanced carefully on HPC systems to enable proper interaction with the host environment. Specifically, MPI implementations must align inside and outside the container, and interactions between the job scheduler and the Program Management Interface (PMIx) must function smoothly. The environment variable \bash{PMIX_SECURITY_MODE=native} was found to be essential and had to be explicitly set in the job's local environment outside of the container. This issue was successfully resolved through close collaboration with system administrators.

On the software level, a race condition in Megatron-LM was discovered, which terminated training runs at seemingly random intervals. Given the link-flipping issues described in \autoref{Sec:Hardware}, this problem was particularly difficult to reproduce and diagnose. It was ultimately mitigated by reducing the number of CPU cores allocated for data loading. 

For training large models not fitting into the GPU memory of a single node, checkpointing required CPU-CPU data transfers. To this end, Infiniband communication had to be enabled in the PyTorch Distributed Gloo backend, as CPU transfers are not possible over NCCL.

Soon after the start of the project, PyTorch started to deprecate its launcher API (\bash{torch.distributed.launch}) we had been using previously. The new API (\bash{torchrun} or \bash{torch.distributed.run}) initially caused issues due to system assumptions that JUWELS Booster did not fulfill. Specifically, it was impossible for PyTorch to correctly identify the used network interface cards (NICs) because they employed non-default hostnames.
We were able to work around these assumptions with extensive analysis and patching of the problematic PyTorch code. The patch was submitted upstream, and a wrapper script was developed to assist users with separately installed PyTorch versions\footnote{\url{https://github.com/HelmholtzAI-FZJ/torchrun_jsc}}.

To overcome Slurm’s runtime limit of \qty{24}{\hour} and to improve robustness against hardware failures, scripts were implemented to support chained job submissions with dependency handling. These scripts automatically resumed training from the latest checkpoint.

Additional challenges arose when building software packages on the login nodes, as recommended for JSC systems. These nodes are intended for tasks such as compiling code and launching jobs and not for running production workloads or intensive computations. However, some software installations assumed they would be built in the same environment in which they would later run, and thus expected access to GPUs. 

\section{Profiling and Visualization}
% different profilers: torch, DLProf, Score-P and NSight System , visualization tools: Tensorboard, LLview
Profiling LLM training on HPC systems is essential to ensure efficient resource utilization and optimize performance. In OpenGPT-X, several profilers were deployed to analyze the performance of the models. Torch Profiler, an integral tool within the PyTorch ecosystem, was used to track performance metrics such as training time, memory consumption, and computational efficiency, helping the team identify and resolve performance bottlenecks. DLProf, NVIDIA's profiler designed for deep learning workloads, provided specialized insights into the performance of models on NVIDIA GPUs. It allows focusing on kernel performance, memory usage, and data movement, crucial for large-scale LLM training. Score-P, a tool supporting parallel applications, was leveraged to profile distributed training on multiple nodes, giving detailed performance data at the node, thread, and function levels. Additionally, NSight Systems by NVIDIA was used to visualize GPU-CPU interactions, memory utilization, and parallel execution.

For analyzing and visualizing the training and system metrics, TensorBoard and LLview~\footnote{\url{https://github.com/FZJ-JSC/LLview}} were used actively. TensorBoard was utilized for visualizing key metrics like loss curves, training accuracy, and hyperparameters, offering real-time dashboards that were crucial for model optimization. 

LLView is a software ecosystem developed at JSC that serves as a highly integrated reporting and visualization tool for jobs running on the HPC infrastructure~\cite{muller24llview}. It enables users to analyze the efficiency with which their allocated resources are being utilised~\cite{maloney2024HPCMonitoring}. In OpenGPT-X it was used to visualize large-scale training metrics across multiple nodes.

By combining these profiling and visualization tools, the OpenGPT-X project achieved more effective training workflows, enabling the team to monitor and optimize the performance of large models and their training infrastructure, ensuring high scalability and resource efficiency throughout the process~\cite{John:1007707}.

\section{Benchmarking and Hyper-parameter optimization}\label{Sec:Benchmark}

% CARAML, energy, model layouting, scaling 
The growing need to identify effective hyperparameter configurations and assess hardware performance in LLM training led to the development of the CARAML benchmark suite~\cite{john2024Caraml} within the OpenGPT-X project. CARAML evaluates AI training workloads by measuring both energy consumption and performance across accelerators by vendors such as NVIDIA, AMD, and Graphcore. The framework supports structured ablation studies and hyperparameter grid search, facilitating the identification of the most efficient configurations when scaling LLM models. This ensures optimized power efficiency and throughput for large-scale AI tasks, enabling effective use of hardware platforms. 

\begin{figure}[!htbp]
  \centering
  \begin{subfigure}{0.9\linewidth}
    \centering
    \includegraphics[width=\linewidth]{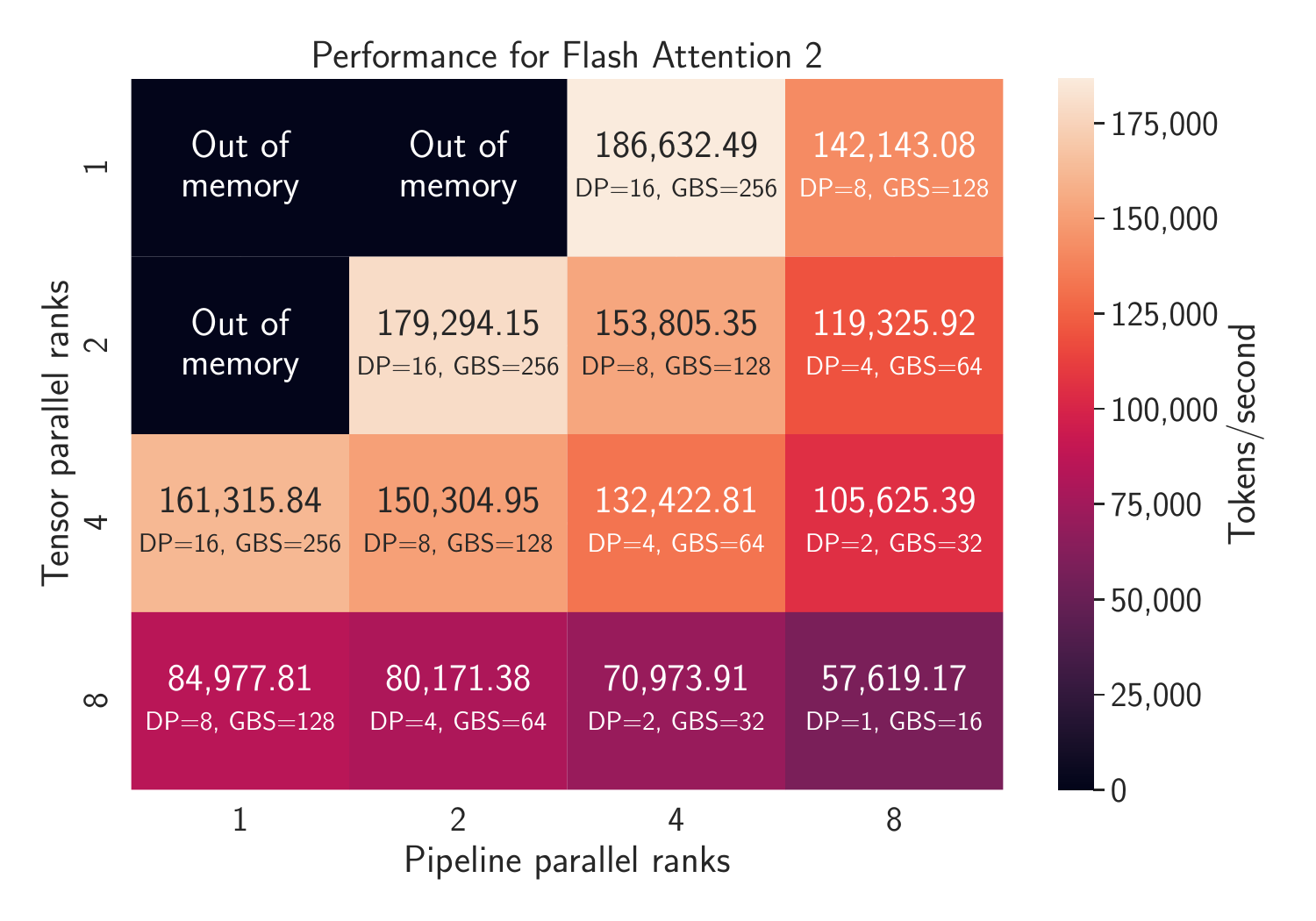}
    \caption{Combinations of tensor and pipeline parallelism and the resulting performance using Flash Attention version 2.4.2.}
    \label{fig:flashattn2}
  \end{subfigure}\par\medskip
  \begin{subfigure}{0.9\linewidth}
    \centering
    \includegraphics[width=\linewidth]{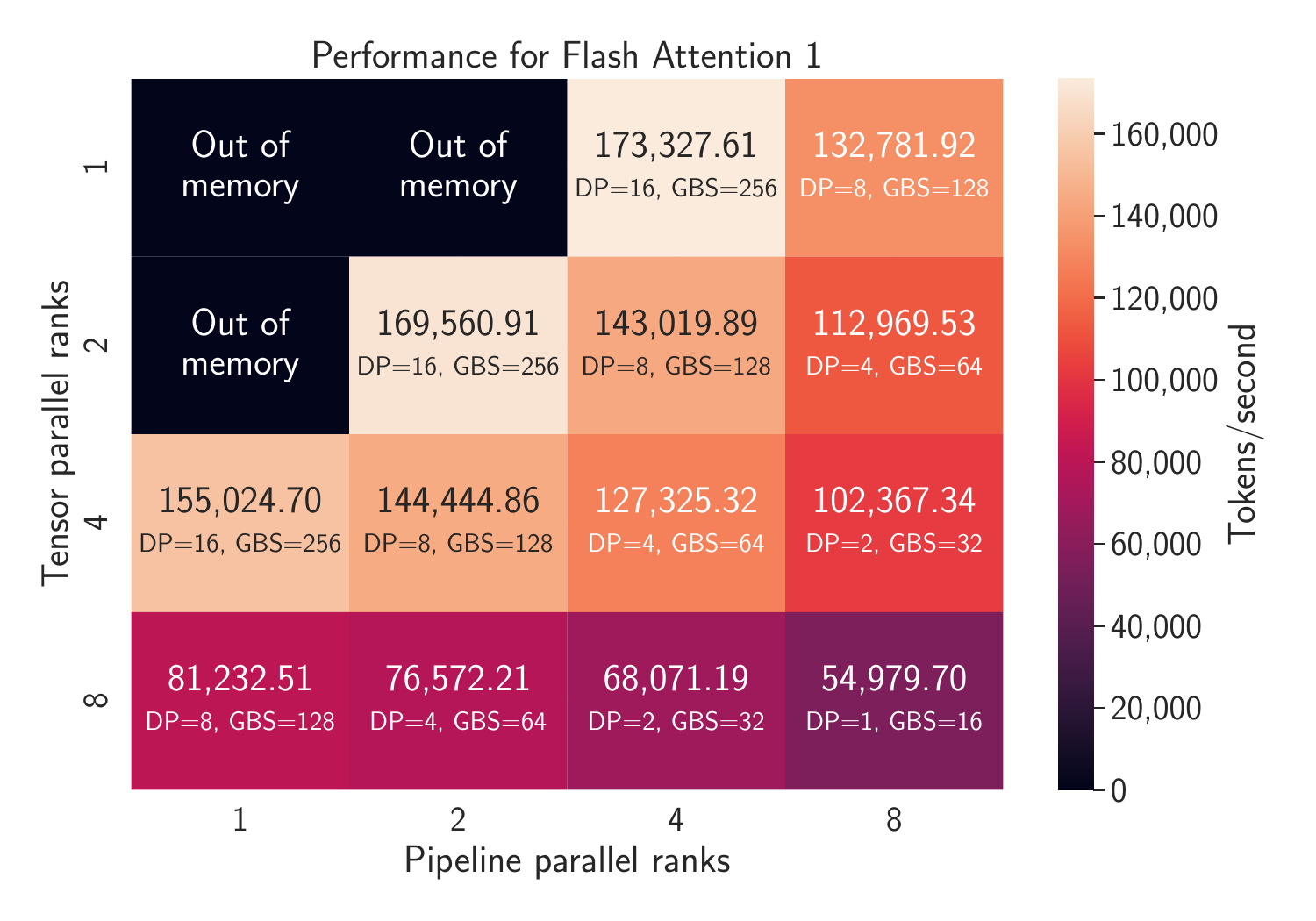}
    \caption{Combinations of tensor and pipeline parallelism and the resulting performance using Flash Attention version 1.0.1.}
    \label{fig:flashattn1}
  \end{subfigure}
  \caption{With a given number of 16 nodes of the JURECA cluster, containing 64 NVIDIA A100 GPUs, the impact of combining tensor and pipeline parallelism on the performance when training a 6.6B parameter model was assesed.}
  \label{fig:heatmaps}
\end{figure}

To illustrate the importance of choosing proper performance parameters, we included throughput results for training a 6.6B parameter model in \autoref{fig:heatmaps}, using Megatron-LM, version 23.06, commit \bash{f772743} and employing the CARAML framwork. Here, 16 accelerated compute nodes of the JURECA HPC system~\cite{Thornig2021jureca} were used, utilizing four A100 GPUs with 40GB device memory each. The final OpenGPT-X models were trained on JUWELS Booster, utilizing the same kind of GPUS while differing in terms of inter- and intra-node interconnects. 

The trained model in these experiments has the same architectural features as the Teuken-7B model~\cite{ali2024teuken7b}, but used a smaller vocabulary size, leading to a slightly lower parameter count. We employed the established paradigm of 3D parallelism, consisting of the orthogonal parallel dimensions determined by tensor parallelism (TP), pipeline parallelism (PP) and data parallelism (DP)~\cite{narayanan2021megatron}. The number of devices, in our test case 64, is given by
\begin{align*}
    \#GPUs = TP \times PP \times DP.
\end{align*}

In data parallelism, each involved device holds a copy of the whole model and processes a local batch of training data. The gradients are synchronized, i.e. averaged, at the end of a backward pass. Because the ranks can work completely in parallel and only one synchronization per step is needed, data parallelism often achieves a higher throughput than tensor or pipeline parallelism. When the model is too large to fit into GPU memory, relying only on data parallelism is not an option, and model parallelism, i.e. tensor or pipeline parallelism, need to be employed. Tensor parallelism distributes weight matrices and activations in the MLP and self-attention blocks of each layer onto the involved devices, requiring synchronization during forward and backward pass at each of these blocks. In pipeline parallelism, the layers are distributed to multiple devices. A batch is divided into micro-batches that can be processed in parallel. Communication is required when sending microbatches to the next pipeline stage, and a pipeline bubble can limit throughput. 

Given a number of devices on a specific system, it is non-trivial to assess the optimal configuration of these parallel ranks to achieve a high throughput. This is why experiments as reported in \autoref{fig:heatmaps} are needed. Sequence parallelism and selective activation recomputation~\cite{korthikanti2022reducingactivation} were turned on. 

The local batch size was kept at 16 to fix its influence on the throughput. The global batch size therefore grew with the number of data parallel ranks (DP). The number of tensor and pipeline parallel ranks were varied and the number of data parallel ranks was inferred as 
\begin{align*}
    DP = \#GPUs / (TP \times PP),
\end{align*}
where the number of GPUs is set to $\#GPUs=64$. 

The experiment was repeated with two distinct versions of flash attention, which constitutes a prominent example of hardware-aware performance optimization. 

The experiments show that it is beneficial for throughput to focus on high data parallelism, as long as the available memory is enough to hold the whole model. On the JURECA system, pipeline parallelism is preferable over tensor parallelism, i.e. with 16 data parallel ranks, the $TP\times PP$ configuration $1\times 4$ leads to a higher throughput than $2\times 2$ and $4 \times 1$. Typically, tensor parallelism is recommended to parallelize within a multi-GPU node, assuming fast NVLink interconnects between GPUs. On JURECA, the lack of these links makes tensor parallelism unattractive, in contrast to systems such as JUWELS Booster. The newer flash attention version improves throughput by up to 8 \%, most prominently in high-throughput settings. Choosing a suboptimal parallelization configuration can have a much more drastic effect than using the older version. The observed effect of the newer version is not as high as reported in the literature~\cite{dao2023flashattention2}.

\begin{figure}[!t]
  \centering
  \begin{subfigure}{0.9\linewidth}
    \centering
    \includegraphics[width=\linewidth]{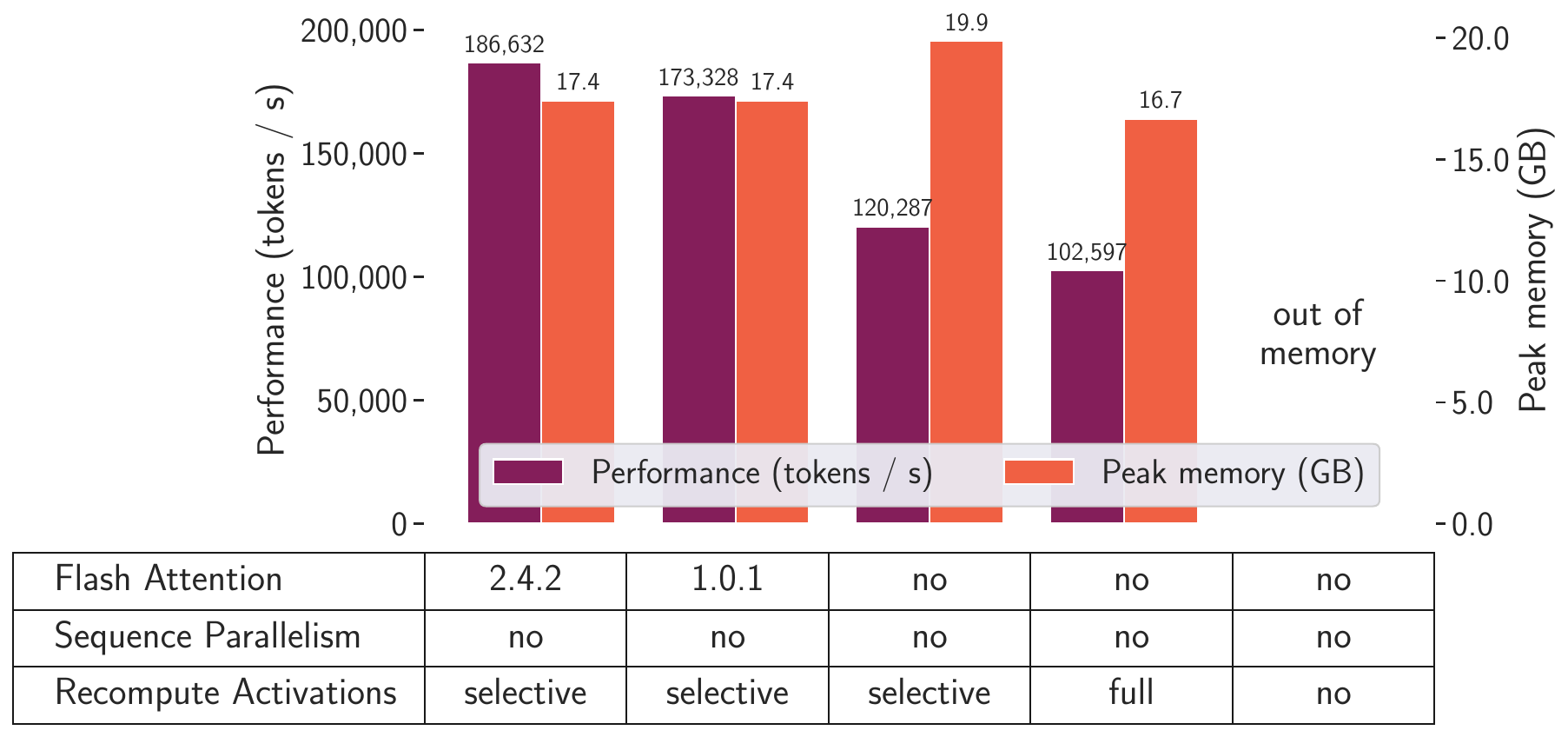}
    \caption{Impact of features for $PP=4$ and $TP=1$.}
    \label{fig:flashattn2}
  \end{subfigure}\par\medskip
  \begin{subfigure}{0.9\linewidth}
    \centering
    \includegraphics[width=\linewidth]{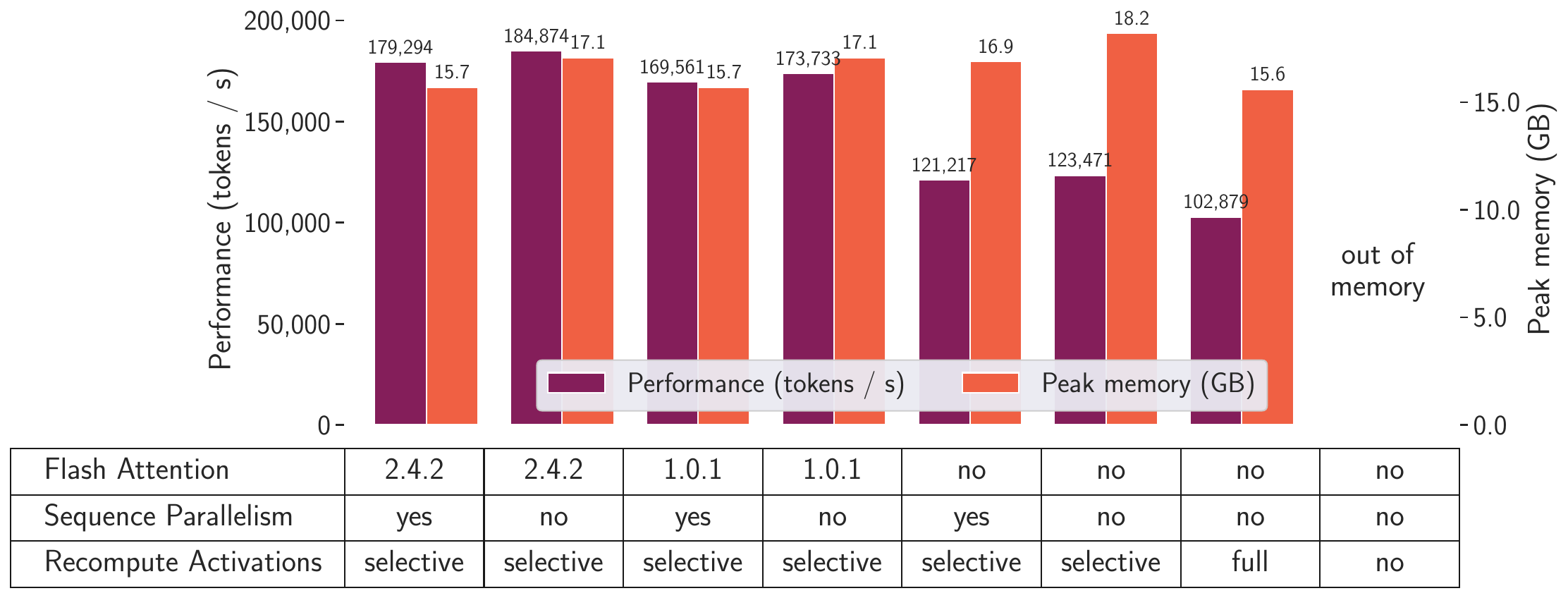}
    \caption{Impact of features for $PP=2$ and $TP=2$.}
    \label{fig:flashattn1}
  \end{subfigure}
  \caption{Further investigations for most performant $TP\times PP$ configurations in \autoref{fig:heatmaps}. The impact of flash attention, sequence parallelism and activation recomputation on performance and memory usage are assessed. }
  \label{fig:featureimpact}
\end{figure}

When training a large language model, further decisions must be made concerning the use certain features, with consequences for performance and memory consumption. To illustrate possible impacts, we performed further experiments for the highest performing configurations of the experiments represented in \autoref{fig:heatmaps}, (a) $TP=1$, $PP=4$ and (b) $TP=2$, $PP=2$. The impact of flash attention in combination with sequence parallelism and activation recomputation~\cite{korthikanti2022reducingactivation} is assessed and results are given in \autoref{fig:featureimpact}. The memory consumption was measured using the \bash{"allocated_bytes.all.peak"} statistics via the dictionary returned by PyTorch's \bash{torch.cuda.memory.memory_stats()}. 

Tensor parallelism distributes operations and activations across the tensor parallel ranks. Sequence parallelism extends the distribution of activations for the dropout and layer normalization layers inside a transformer block, that are not covered by tensor parallelism alone. As these operations are performed on each token independently, the distribution can happen along the sequence dimension. This way, no device ever needs to store all activations at once, reducing the peak amount of memory needed per device.

As the distribution happens along the tensor parallel dimension, sequence parallelism can not be employed for the $TP=1$ configuration. For $TP=2$, $PP=2$, a reduction in memory requirements can be observed at the expense of performance.  

In order to keep memory requirements low, activation recomputation is essential. In our setup, employing GPUs with 40GB of memory, disabling this feature leads to out of memory errors. With flash attention, selective activation recomputation is enabled automatically. In the setting without flash attention, full activation recomputation further reduces memory requirements at the cost of performance. For Megatron-LM, the interplay of various options (\bash{--use-flash-attn},  \bash{--recompute-activations}, \bash{--recompute-granularity}, \bash{--recompute-method}, \bash{--recompute-num-layers}) is not obvious and some options overwrite others. 

The systems of the JURECA Evaluation Platform (see \autoref{Sec:JurEval}) were benchmarked using CARAML~\cite{john2024Caraml}, with detailed measurements of energy consumption and training performance (see Fig.2 in~\cite{john2024Caraml}).

The OpenGPT-X training setup, based on Megatron-LM, was integrated into the JUPITER benchmark suite~\cite{Herten_2024}. This suite consists of representative workloads used during the procurement process of JUPITER, Europe’s first exascale HPC system, expected to become operational in 2025. The publicly available, reproducible benchmark repository~\footnote{https://gitlab.jsc.fz-juelich.de/jubench/jubench-ad/-/blob/main/Megatron-LM/DESCRIPTION.md}, provides a valuable resource for researchers seeking to run large-scale LLM training workloads on HPC clusters. 

The goal of the benchmarking efforts during procurement was to test the potential future system's scaling, communication and computation capacities for training large, transformer-based neural networks (see Fig. 2 in~\cite{Herten_2024}). Results demonstrated that the NVIDIA GH200 significantly outperformed the A100 GPUs, offering approximately twice the performance. 

Additional extensive benchmarks regarding LLM training capacities were conducted on the JEDI system (see \autoref{Sec:JurEval}), further validating the expected AI training capabilities of the upcoming JUPITER system.

\section{Conclusions}
\label{Sec:Conclusions}
The OpenGPT-X project provided valuable insights into the challenges and best practices of training large-scale language models on HPC infrastructure. We conclude with a set of concise takeaways that may guide similar future efforts:

\begin{enumerate}
\item \textbf{Build on existing work:} Starting from well-established, community-maintained codebases significantly accelerates progress and avoids unnecessary duplication of effort.
\item \textbf{Keep up with new developments but stay focused:} Staying up-to-date with rapid developments in the field is essential, but efforts should be grounded in clearly defined project goals to maintain direction and deliver results.
\item \textbf{Acknowledge the engineering effort:} Training LLMs at scale is not only a research challenge but also an extensive engineering endeavor that demands careful system integration and infrastructure adaptation.
\item \textbf{Understand your system:} A deep understanding of the underlying hardware and software stack, including its limitations and quirks, is critical for achieving reliable and efficient training runs.
\item \textbf{Collaborate with system administrators:} Close communication with HPC support staff helps resolve system-specific issues early and improves overall workflow stability and performance.
\item \textbf{Document and communicate:} Clear documentation, reproducible setups, and shared knowledge bases are essential for onboarding, collaboration, and long-term project sustainability.
\end{enumerate}

These lessons contribute to a growing body of practical knowledge on LLM training at scale and can inform both academic and industrial efforts in the evolving AI infrastructure landscape.

\section*{Acknowledgment}
This work was funded by the German Federal Ministry for Economic Affairs and Climate Action (BMWK) through the project OpenGPT-X (project no. 68GX21007D).

Work presented here made extensive use of the JSC clusters JUWELS Booster, JUWELS Cluster, JURECA-DC, the JURECA-DC Evaluation Platform, the WestAI infrastructure, and the JUPITER enablement platform JEDI, which we greatly acknowledge.

\printbibliography[heading=bibintoc]

\appendix
\section{Appendix}\label{LLMSbatch}
Example Slurm job script to launch Megatron-LM training on JUWELS booster using NGC containers\footnote{\url{https://catalog.ngc.nvidia.com/}}.

\begin{minted}{bash}
#!/bin/bash
#SBATCH --job-name=800M_model
#SBATCH --account=account
#SBATCH --partition=booster
#SBATCH --nodes=2
#SBATCH --gres=gpu:4                    # numbe of GPUs
#SBATCH --ntasks-per-node=1             # Crucial - only one task per dist per node !
#SBATCH --cpus-per-task=48              # Slurm 22.05: srun doesnot inherit this variable from sbatch
#SBATCH --time=00:20:00                 # maximum execution time (HH:MM:SS)
#SBATCH --threads-per-core=1            # using only real cores, no SMT
#SBATCH --output=%x-%j.out              # output file name
#SBATCH --error=%x-%j.err               # error file name



# explicitly setting srun environment variable to inherit from SBATCH
export SRUN_CPUS_PER_TASK=${SLURM_CPUS_PER_TASK}

# Enable logging
set -euo pipefail
set -x

echo "START TIME: $(date)"
#### Input data ####
# Data: https://huggingface.co/bigscience/misc-test-data/resolve/main/stas/oscar-1GB.jsonl.xz
# Tokeniser Vocab: 
# https://s3.amazonaws.com/models.huggingface.co/bert/gpt2-vocab.json
# Tokenizer Merge file: 
# https://s3.amazonaws.com/models.huggingface.co/bert/gpt2-merges.txt
VOCAB_FILE=gpt2-vocab.json
MERGE_FILE=gpt2-merges.txt
DATA_PATH=oscar/oscar_text_document


#### Output paths ####
DATA_OUTPUT_PATH="${SLURM_JOB_ID}_${SLURM_JOB_NAME}"
CHECKPOINT_PATH="$DATA_OUTPUT_PATH"/checkpoints
TENSORBOARD_PATH="$DATA_OUTPUT_PATH"/tensorboard
CODECARBON_PATH="$DATA_OUTPUT_PATH"/codecarbon
CACHE_DIR="$DATA_OUTPUT_PATH/.cache"
LOGS_PATH="$DATA_OUTPUT_PATH"/logs
TORCHELASTIC_ERROR_FILE=$LOGS_PATH/torch_distribute_error.txt

mkdir -p $LOGS_PATH

# copy this batch script into log directory for reproducibility
if [ -e "$0" ]; then
    cp -p "$0" "$LOGS_PATH/batch-${SLURM_JOB_NAME}-${SLURM_JOB_ID}.sh"
fi

#### Environment variables ####
export LOAD_CHECKPOINTS=false
export HF_DATASETS_OFFLINE=1
export TRANSFORMERS_OFFLINE=1
export CUDA_DEVICE_MAX_CONNECTIONS=1

export CXX=g++
export CC=gcc
# force crashing on nccl issues like hanging broadcast
export NCCL_ASYNC_ERROR_HANDLING=1
# handle timeouts
export NCCL_IB_TIMEOUT=50
export UCX_RC_TIMEOUT=4s
export NCCL_IB_RETRY_CNT=10
# setting IB for out of band communication
export NCCL_SOCKET_IFNAME=ib0
export GLOO_SOCKET_IFNAME=ib0
# For debugging 
export LOGLEVEL=INFO 
# export CUDA_LAUNCH_BLOCKING=1
# export NCCL_DEBUG=INFO
# export NCCL_DEBUG_SUBSYS=ALL
# export TORCH_DISTRIBUTED_DEBUG=INFO


##### Network parameters #####
MASTER_ADDR=$(scontrol show hostnames $SLURM_JOB_NODELIST | head -n 1)
# Allow communication over InfiniBand cells.
MASTER_ADDR="${MASTER_ADDR}i"
# Get IP for hostname.
MASTER_ADDR="$(nslookup "$MASTER_ADDR" | grep -oP '(?<=Address: ).*')"
MASTER_PORT=6000


cd $MEGATRON_LM_REPO  # Megatron-LM: https://github.com/NVIDIA/Megatron-LM
CLEAN_PREV_JIT_BUILD=0
rm -f megatron/fused_kernels/build/lock
# Not rebuilding fused kernels have led to errors, so they are
# optionally deleted.
((CLEAN_PREV_JIT_BUILD)) && rm -rf megatron/fused_kernels/{build,__pycache__}

MODEL_SIZE=800M
##### Parallel model layouting #####
# This is an example configuration, not an optimized one
GPUS_PER_NODE=4         # Minimum required number of GPUs: PP_SIZE * TP_SIZE 
NNODES=$SLURM_JOB_NUM_NODES
PP_SIZE=1               # NLAYERS must be a multiple of PP_SIZE here
TP_SIZE=1                # TP_SIZE <= GPUS_PER_NODE (preferred)
MICRO_BATCH_SIZE=4
GLOBAL_BATCH_SIZE=512

#### Notes ####
# NGPUS = NNODES * GPUS_PER_NODE = TP_SIZE * PP_SIZE * DP_SIZE.
# Given NNODES, GPUS_PER_NODE, PP_SIZE and TP_SIZE. DP_SIZE is
# inferred automatically as NGPUS/(PP_SIZE * TP_SIZE).
# GLOBAL_BATCH_SIZE has to be divisible by MICRO_BATCH_SIZE*DP_size

#### Hyperparameters ####
NLAYERS=16
NHIDDEN=2048
NHEADS=8
SEQ_LEN=2048
VOCAB_SIZE=50257

SAVE_INTERVAL=3000
LOG_INTERVAL=10
EVAL_INTERVAL=40000

TRAIN_SAMPLES=244_140
TRAIN_TOKENS=500_000_000

LR_DECAY_SAMPLES=126_953_125
LR_WARMUP_SAMPLES=183_105

OPTIMIZER_ARGS=" \
    --optimizer adam \
    --adam-beta1 0.9 \
    --adam-beta2 0.95 \
    --adam-eps 1e-8 \
    --lr 0.00025 \
    --min-lr 0.000025 \
    --lr-decay-style cosine \
    --lr-decay-samples $LR_DECAY_SAMPLES \
    --lr-warmup-samples $LR_WARMUP_SAMPLES \
    --clip-grad 1.0 \
    --weight-decay 1e-1 \
    --use-distributed-optimizer \
    "

EXIT_OPTS=" \
    --exit-duration-in-mins 60 \
    "
    	
GPT_ARGS=" \
    --num-layers $NLAYERS \
    --hidden-size $NHIDDEN \
    --num-attention-heads $NHEADS \
    --seq-length $SEQ_LEN \
    --max-position-embeddings $SEQ_LEN \
    --micro-batch-size $MICRO_BATCH_SIZE \
    --global-batch-size $GLOBAL_BATCH_SIZE \
    --train-samples $TRAIN_SAMPLES \
    --vocab-file $VOCAB_FILE \
    --merge-file $MERGE_FILE \
    --bf16 \
    --seed 42 \
    --recompute-activations \
    --init-method-std 0.0048 \
    --position-embedding-type rope \
    --use-flash-attn \
    --sequence-parallel \
    $OPTIMIZER_ARGS \
    $EXIT_OPTS \
    "
  
OUTPUT_ARGS=" \
    --log-interval $LOG_INTERVAL \
    --save-interval $SAVE_INTERVAL \
    --eval-interval $EVAL_INTERVAL \
    --eval-iters 10 \
    --tensorboard-dir $TENSORBOARD_PATH \
    --tensorboard-queue-size 5 \
    --log-timers-to-tensorboard \
    --log-batch-size-to-tensorboard \
    --log-validation-ppl-to-tensorboard \
    "

export LAUNCHER="python -u -m torch.distributed.run \
    --nproc_per_node $GPUS_PER_NODE \
    --nnodes $NNODES \
    --rdzv_endpoint $MASTER_ADDR:$MASTER_PORT \
    --rdzv_backend c10d \
    --max_restarts 0 \
    --tee 3 \
    "
    
export CMD=" \
    $(pwd)/pretrain_gpt.py \
    --tensor-model-parallel-size $TP_SIZE \
    --pipeline-model-parallel-size $PP_SIZE \
    $GPT_ARGS \
    $OUTPUT_ARGS \
    --save $CHECKPOINT_PATH \
    --data-path $DATA_PATH \
    --data-impl mmap \
    --split 949,50,1 \
    --distributed-backend nccl \
    "
    
if [ "$LOAD_CHECKPOINTS" = true ] ; then
    export CMD="$CMD\
        --load $CHECKPOINT_PATH \
        "
fi
echo $CMD
# Pull and build a conatiner from 
# https://catalog.ngc.nvidia.com/orgs/nvidia/containers/pytorch/tags

SINGULARITY_FILE=ngc_torch.sif
srun --jobid $SLURM_JOBID --cpu-bind=v --mpi=pmi2 \
    apptainer exec --bind="$MEGATRON_LM_REPO",data,"$DATA_OUTPUT_PATH" \
    --nv "$SINGULARITY_FILE" \
    bash -c "$LAUNCHER  $CMD" 2>&1 | tee -a "$LOGS_PATH"/main_log.txt) &

echo "END TIME: $(date)"
\end{minted}

\end{document}